\documentclass[prb,twocolumn,amsmath,amssymb]{revtex4}

\usepackage{graphicx}% Include figure files
\usepackage{dcolumn}% Align table columns on decimal point
\usepackage{bm}% bold math

\begin{document}

\title{Graphene-based spin-pumping transistor}% Force line breaks with \\

\author{F. S. M. Guimar\~aes$^a$, A. T. Costa$^{a}$, R. B. Muniz$^{a}$ and M. S. Ferreira$^{b}$\footnote{E-mail address: ferreirm@tcd.ie  (M. S. Ferreira)    }} 
\affiliation{
(a) Instituto de F\'{\i}sica, Universidade Federal Fluminense, Niter\'oi, Brazil \\
(b) School of Physics, Trinity College Dublin, Dublin 2, Ireland }

\date{\today}% It is always \today, today,
             %  but any date may be explicitly specified

\begin{abstract}
We demonstrate with a fully quantum-mechanical approach that graphene can function as gate-controllable transistors for pumped spin currents, {\it i.e.}, a stream of angular momentum induced by the precession of adjacent magnetizations, which exists in the absence of net charge currents. Furthermore, we propose as a proof of concept how these spin currents can be modulated by an electrostatic gate. Because our proposal involves nano-sized systems that function with very high speeds and in the absence of any applied bias, it is potentially useful for the development of transistors capable of combining large processing speeds, enhanced integration and extremely low power consumption. 
\end{abstract}

% PACS, the Physics and Astronomy
                 
\maketitle
\bibliographystyle{abbrv} % Choose Phys. Rev. style for bibliography

The field of spintronics, responsible for a number of exciting proposals for future electronic devices, is based on ingenious ways of controlling not only how the electron current flows across magnetic (M) and non-magnetic (NM) hybrid materials, but also how the electron spin propagates throughout this environment \cite{spintronics-review}. One of the most sought-after applications in the field is the generation of efficient spin transistors capable of modulating the transport properties, depending on the spin of the current-carrying electrons \cite{datta}. Such a spin-polarized current has been produced in several instances, detected in a variety of ways \cite{examples} and has as a common characteristic the fact that the scattering contrast that exists between M and NM regions in these hybrid materials leads to the sought transistor behavior. 

Current challenges for producing spin transistors are in designing structures in which: ({\it i}) the spin coherence length is larger than or at least comparable with the system size;  ({\it ii}) spin-polarized currents can be efficiently injected from the M to the NM regions of the system. Here we argue that graphene has the ideal properties to overcome both limitations and propose that it can be used as a spin transistor that allows control of spin transport through an electrostatic gate. Furthermore, our proposal involves nanoscaled systems that function with very high speeds and in the absence of any applied bias, which can in principle pave the way towards the construction of transistors combining increasingly large data processing speeds, enhanced integration and extremely low electric power consumption.  

The first bottleneck is likely to be overcome with the use of graphene if we recall numerous predictions that have been made for extremely long spin-coherence lengths for these materials \cite{yazyev, kane, peres, wimmer}. In fact, recent experimental evidences confirming these predictions have also indicated how much room there is for increasing this length \cite{lundeberg}. The second obstacle arises primarily due to the inherent contact resistance that exists between the M and NM regions. It has a spin-independent part, commonly referred to as background resistance, added to a contribution that depends on the electron spin. Depending on the value of this background resistance, the conductance contrast between the different spin channels may be relatively very small, leading to a considerable reduction in the spin polarization of the injected current.  

As opposed to a spin-polarized charge current, pumped spin currents consisting of angular momentum flow without the necessity of a net charge current may be the answer to the spin-injection limitation. In fact, Tserkovnyak {\it et al} proposed a mechanism for pumping spin current into NM metals through the precession of an adjacent magnetization \cite{Tser, bauer-e-cia}. In this case, angular momentum from the moving magnetization is transferred to the conduction electrons, creating a spin disturbance that propagates throughout the metallic conduit. This spin flow is produced without an applied voltage, involves no net electrical current, and may be used to excite other magnetic units also in contact with the NM metal.

Furthermore, we have recently demonstrated that carbon nanotubes are capable of carrying spin currents generated by precessing magnetic moments for long distances with minor dispersion and with adjustable degrees of attenuation. We have also shown that these magnetic excitations travel with the nanotube Fermi velocity and are capable of setting other spins into precession even when they are a long distance apart \cite{submitted}. Because of the resemblance between their electronic dispersion relations, very similar behaviour is expected for graphene.

For meeting the aforementioned challenges, graphene seems the ideal candidate for the NM component of spin transistors. Moreover, the possibility of developing magnetic moments when graphene is doped with transition metal atoms \cite{arkady-prl, vene-prb, gan, ushiro} or when in contact to ferromagnetic metals \cite{jpcm04, prb04} indicates that it is relatively straightforward to create a M region within graphene \cite{esquinazi}. The goal of the present paper is therefore to demonstrate that the spin current generated by the precessing magnetization of the M region in contact to graphene can be modulated when propagating across an electrostatic gate. In other words, we propose that graphene can be used as an efficient gate-controllable spin-pumping transistor. More than a mere proof of concept, here we describe the dynamic response of the spin disturbance taking full account of the electronic structure of the system. Our approach discloses some interesting quantum aspects of the spin dynamics which are concealed by semiclassical treatments. It combines the spin pumping characteristics with the suitable features of graphene as good conduits for magnetic information.

Let us start by defining the system to be considered. As in typical spin-valve transistors, two regions (M and NM) are needed. Here we consider as our NM part an infinitely long nanoribbon of graphene of width $W = 8$ atoms. Because our results are not crucially dependent on the ribbon shape, we opt to work with armchair-shaped edges. There are numerous possibilities for the other region, the most obvious of which is to promote the interaction of graphene with some magnetic object. Magnetic atoms substitutionally inserted into the hexagonal atomic structure of graphene are known to display sizable magnetic moments \cite{arkady-prl, vene-prb} and will be used here to represent the M part of the system. Other objects such as magnetic nanoparticles or ferromagnetic substrates can also be used but substitutional atoms are by far the simplest and yet capture the essence of the phenomenon we intend to describe. Four different setups are considered, all of which illustrated in the insets of Figure 1: (a) one single magnetic atom placed at the centre of the ribbon; (b) two identical atoms placed a distance $D$ apart; (c) one stripe of magnetic atoms placed across the ribbon width; (d) one stripe of magnetic atoms with another isolated atom a distance $D$ apart. While a pristine stripe of substitutional impurities is unlikely to exist, here it serves the purpose of testing how the transistor behaviour responds to the size of the magnetic material responsible for generating the spin current. 

As previously mentioned, the pumped spin current is generated by the precession of a magnetization in contact with a metallic medium. We assume that the magnetization is originally in equilibrium, pointing along an arbitrary z-direction, and that it is set into precession by a time-dependent transverse field $h_\perp (t)$. In practical terms, there are different ways of producing this time-dependent perturbation \cite{bauer-e-cia,bauer03,wees}. To determine how graphene transports spin currents, we must assess how such a localized magnetic excitation propagates across the structure. In other words, we must investigate how this excitation disturbs the spin balance of the system not only where the magnetic atoms are located but also, and more importantly, how the local spin dynamics is affected within the NM graphene. This is naturally manifested by the spin susceptibility $\chi$, which reflects how the spin degrees of freedom of a system respond to a magnetic excitation. 

To calculate the spin susceptibility one needs the Hamiltonian describing the electronic structure of the unperturbed system, which we assume is given by  $\hat{H} =   \sum_{i,j,\sigma} \gamma_{ij} \, \ {\hat
c}_{i\sigma}^\dag \, {\hat c}_{j\sigma} + \sum_{\ell,\sigma}  {U \over 2}  \, {\hat n}_{\ell \sigma} \, {\hat n}_{\ell {\bar \sigma}} + \hat{H}_Z$. Here, $\gamma_{ij} $ represents the electron hopping between nearest neighbor sites $i$ and $j$,  $\hat{c}_{i\sigma}^{\dag}$ creates an electron with spin $\sigma$ at site $i$, the sum in $\ell$ is over the sites occupied by magnetic atoms, $\hat{n}_{\ell \sigma} = \hat{c}_{\ell \sigma}^{\dag} \hat{c}_{\ell \sigma}$ is the corresponding electronic occupation number operator, and $U$ represents an effective on-site interaction
between electrons in the magnetic sites, which is neglected elsewhere. Finally, $H_Z$ plays the role of a local Zeeman interaction that defines the $\hat{z}$-axis as the equilibrium direction of the magnetization. The
Hamiltonian parameters can be obtained from density-functional-theory calculations so that the electronic structure of the doped system is well described \cite{charlier-prl04, rocha}. Although the presented results are for Mn atoms, other substitutional magnetic impurities may be employed \cite{arkady-prl}. In our calculations we fix the Fermi energy $E_{F}=0$, and use $\gamma_{i,j} = 2.7$ eV for the nearest-neighbour hopping terms. We take the number of $d$-electrons in the Mn sites $n_m = 1$, $U=20$ eV, and assume a local Zeeman energy splitting of 1 meV \cite{obs}.  Spin-orbit coupling is neglected for being very small compared to the other relevant energy scales. 

The time-dependent transverse spin susceptibility is defined as $\chi_{m,\ell}(t) =
-{i \over \hbar} \Theta(t)\langle[{\hat S}_m^+(t),{\hat S}_\ell^-(0)]\rangle$,
where $\Theta(x)$ is the heaviside step function, and ${\hat S}_m^+$ and ${\hat
S}_m^-$ are the spin raising and lowering operators at site $m$, respectively.
The indices $\ell$ and $m$ refer to the locations where the field is applied and
where the response is measured, respectively. In our case, we induce a precession of the
magnetic moments on sites $\ell$ and we wish to observe the resulting spin
disturbance at an arbitrary site $m$. This response is fully described by
$\sum_\ell \chi_{m,\ell}(t)$. Within the random phase approximation, this susceptibility may
be calculated in frequency domain, and in matrix form it is given by
$\chi(\omega) = [1 + \chi^0(\omega)\, U]^{-1} \, \chi^0(\omega)$, 
where $\chi^0$ is the Hartree-Fock susceptibility.

Recently we have investigated the spin disturbance in carbon nanotubes as a function of time and have identified that pulsed magnetic excitations travel across these materials with their corresponding Fermi velocities, with very little deformation and with tunable degrees of attenuation, something that has been attributed to the distinctive linear dispersion relation displayed by nanotubes \cite{submitted}. Graphene has a similar linearity in their dispersion relation and behaves in exactly the same way. However, rather than studying the time-resolved spin excitations, in this paper we focus on the frequency-resolved response. The key quantity for us is $\vert \chi_{m,\ell}(\omega)\vert$ because it is proportional to the amplitude of the spin disturbance at site $m$ due to a time-dependent transverse magnetic field applied at site $\ell$. In other words, it corresponds to the precession amplitude acquired by the electron spin localized at a an arbitrary site $m$ due to a harmonic perturbation of frequency $\omega$ produced at the magnetic site $\ell$. Unsurprisingly, when investigated as a function of frequency this quantity displays a very distinctive peak close to its Larmour-frequency resonance. Without any loss of generality, in what follows we use this value as our choice of excitation frequency.

\begin{figure}
\includegraphics[width = 8.5cm] {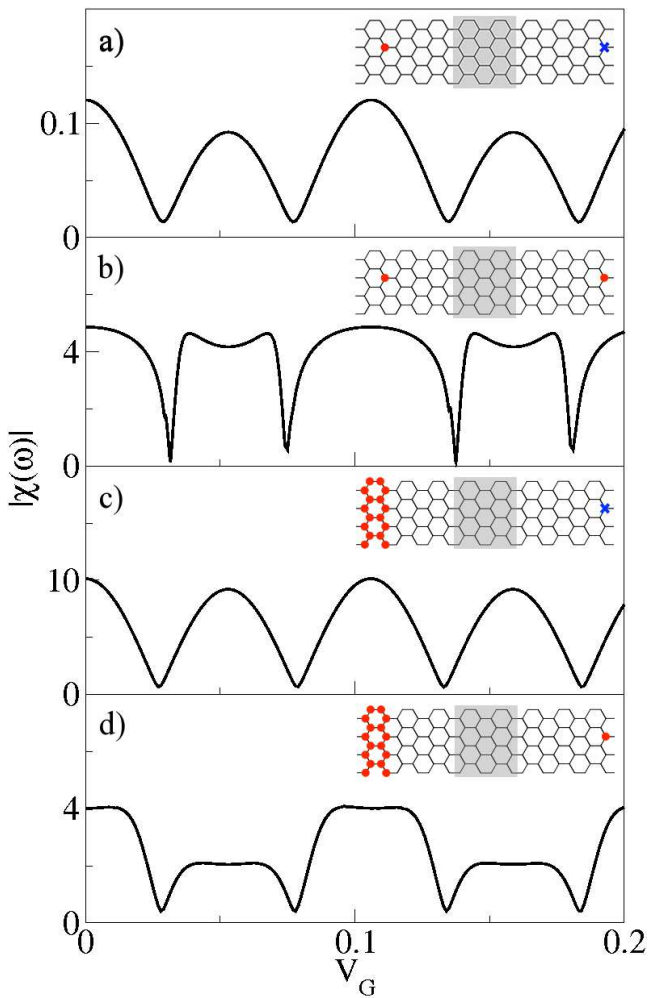}
\caption{Spin disturbance, represented by the spin susceptbility $\vert \chi_{m,\ell}(\omega)\vert$ (in units of $\hbar/{\rm eV}$), as a function of the gate voltage $V_g$ (in units of eV). Four different setups are considered: (a) one single magnetic atom; (b) two identical atoms separated by a distance $D$; (c) one stripe of magnetic atoms placed across the ribbon width; (d) one stripe of magnetic atoms with another isolated atom a distance $D$ apart. The insets depict schematic representations of each one of these setups, where the red circles represent the magnetic atoms, the blue cross shows the position where the spin disturbance is probed even in the absence of a magnetic object, and the gray area in the middle indicates the gated region under the action of $V_g$. All spin disturbances are probed at the marked sites on the right-hand side of the insets, either as a red circle or as a blue cross. For the sake of clarity, results for panels (b) and (d) have been divided by $10^2$ and $10^4$, respectively.}
\label{figure_1}
\end{figure}

The fact that the electron spin probed at arbitrary graphene sites is affected by a perturbation acting where the magnetic atoms are located indicates the existence of a pumped spin current traveling through graphene. The ease with which graphene is electrostatically gated \cite{ah-review} suggests that one may modulate the spin current pumped across these materials. We mimic the effect of a gate voltage by introducing a region of length $L$ separating the magnetic sites $\ell$ and the location where we probe the spin disturbance, in which the electrostatic potential is $V_g$. Figure 1 shows results for the spin disturbance as a function of $V_g$ for all four different setups. Results for $V_g<0$ are not shown but the spin disturbance is even with respect to $V_g$. It is instructive to start by commenting on the results for $V_g = 0$. Sizable values are obtained in all four cases, meaning that graphene, similarly to nanotubes, is indeed a good conduit for the spin current \cite{submitted}. The big difference in magnitude results from the fact that the amplitude of the spin disturbance is largely amplified by the presence of local magnetic moments at the probing site.  

Being a good conduit for the current is necessary but not sufficient to generate efficient transistors. In addition, one needs a way of switching the current between on and off states, something often achievable with a varying gate potential. As a function of $V_g$, all cases show oscillations in $\vert \chi_{m,\ell}(\omega)\vert$ alternating between sizable (on) and vanishingly small (off) values with the same periodicity of $\Delta = 0.106$ eV. This means that a small variation of $V_g$ is capable of controlling whether or not a magnetic object can be excited by a remotely induced precession.  

To map how the spin disturbances are spatially distributed, it is worth investigating their position dependence. For a fixed value of $V_g$ in the case depicted by the inset of figure \ref{figure_1}(c), we assess how the pumped spin current is affecting the spin balance along its path by plotting $\vert \chi_{m,\ell}(\omega)\vert$ as a function of the probing position $m$. Figure 2 displays such a plot where the white (gray) background represents the non-gated (gated) region. The top panel shows results for three different values of $V_g = 0, \Delta, 2 \Delta$, all corresponding to peaks in Figure 1(c). The spin disturbance outside the gated area is indistinguishable in all three cases, confirming the full transmission across the gate. The distinction appears only inside the gate where oscillatory features arise as we increase the corresponding values of $V_g$. A monotonic form is found for the black line representing $V_g=0$, due to the familiar commensurability effect that exists in half-filled electronic bands in hexagonal atomic structures \cite{coupling1,suppression,carbon,rkky}. As a matter of fact, for the same reason, this monotonicity is also found everywhere outside the gated region. As the gate voltage is increased to $V_g = \Delta$ and  $V_g = 2 \Delta$  we depart from a commensurate regime and enter a region where clear oscillatory patterns are seen. Despite the changes in $V_g$, the overall effect on the transmission is identical to the case $V_g = 0$ because the oscillations fit exactly into the length $L$ of the gated region with an even number of half-wavelengths. 

In the case of vanishingly small spin disturbances, shown in the bottom panel of figure 2, the values of $V_g$ for which $\vert \chi_{m,\ell}(\omega)\vert$ is minimal depend on the actual setup considered. Using the setup shown in the inset of figure 1(c), we find that  $V_g^0 = 0.027 \, {\rm eV}$ induces a monotonic decay of the spin disturbance towards very small values, meaning that the pumped spin current has been blocked and is unable to proceed beyond the gated area. While there should be no commensurate effect for $V_g \neq 0$, small values of $V_g$ induce oscillations that are too long to be seen within the length $L$. Subsequent values of $V_g =  V_g^0 + \Delta$ and $V_g =  V_g^0 + 2 \Delta$ shorten the oscillation periods and once again appear as oscillations that fit the length $L$, this time however, with an odd number of half-wavelengths inducing vanishingly small values for $\vert \chi_{m,\ell}(\omega)\vert$.

\begin{figure}
\includegraphics[width = 8.5cm] {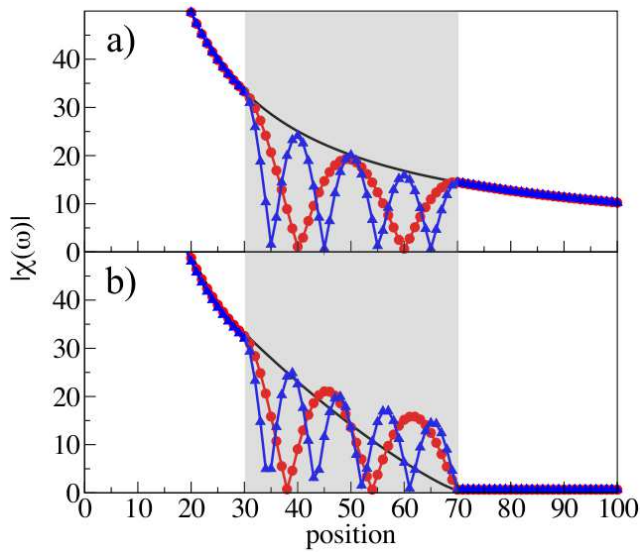}
\caption{Spin disturbance, represented by the spin susceptbility $\vert \chi_{m,\ell}(\omega)\vert$ (in units of $\hbar/{\rm eV}$), as a function of the probing position for different values of $V_g$. All results are for the case of one stripe of magnetic atoms placed across the ribbon width (panel (c) of figure 1). The gray area represents the gated region. (a) $V_g = 0$ (black line); $V_g = 0.106 \, {\rm eV}$ (red line with circles);  $V_g = 0.212 \, {\rm eV}$ (blue line with triangles). (b) $V_g = 0.027 \, {\rm eV}$ (black line); $V_g = 0.133 \, {\rm eV}$ (red line with circles);  $V_g = 0.239 \, {\rm eV}$ (blue line with triangles).  }
\label{figure_2}
\end{figure}

The results displayed above can be easily interpreted in terms of an interference pattern created by the gate. Although pumped spin currents travel without the necessity of net charge currents, they are still carried by electrons. The effect of a gate is to change the characteristic wavelength with which electrons travel inside that region. Depending on how this characteristic wavelength compares to the gate length $L$, one may have constructive or destructive interference, leading to a full transmission or the complete reflection of the pumped spin current. This is a clear signature of quantum interference and is directly analogous to a Fabry-Perot interferometer which possesses certain resonance energies for total transmission together with others where traveling is entirely blocked. 

In summary, we have shown with a full quantum mechanical approach that spin current can be pumped into graphene when it is in contact to a M region whose magnetization is set into precession motion. This spin current requires no applied bias to be pumped across graphene and does not depend on any net charge current. In addition, the pumped current is perfectly controllable with the introduction of a simple gate that modulates the transmission as a result of quantum interference. Finally, the ease with which graphene is gated and the facility with which current state-of-the-art experimental techniques are used to produce spin-pumping devices suggests that the proposal described here is of simple implementation. The ideas presented here may be valuable in overcoming some of the limitations displayed by conventional devices and will motivate new attempts to produce spin transistors with large processing speeds, enhanced integration and extremely low power consumption.


\begin{references}
\bibitem{spintronics-review} I. Zutic, J. Fabian and S. Das Sarma, Rev. Mod. Phys. {\bf 76}, 323 (2004); D. D. Awschalom and M. E. Flatte, Nat. Phys. {\bf 3}, 153 (2007)
\bibitem{datta} S. Datta and Das, Appl. Phys. Lett. {\bf 56}, 665 (1990)
\bibitem{examples} R. Lu, Hai-Zhou Lu, Xi Dai and J. Hu, J. Phys.: Condens. Matt. {\bf 21}, 495304 (2009); H. Jang and I. Appelbaum, Phys. Rev. Lett. {\bf 103}, 117202 (2009); V. S. Pribiag , I. N. Krivorotov , G. D. Fuchs , P. M. Braganca , O. Ozatay , J. C. Sankey , D. C. Ralph and R. A. Buhrman, Nat. Phys. {\bf 3}, 498 (2007) 
\bibitem{yazyev} O. V. Yazyev, Nano Lett. {\bf 8}, 1011 (2008) 
\bibitem{kane} C. L. Kane and E. J. Mele, Phys. Rev. Lett. {\bf 95}, 226801 (2005) 
\bibitem{peres} N. M. R. Peres, F. Guinea and A. H. Castro Neto, Phys. Rev. B {\bf 72}, 174406 (2005) 
\bibitem{wimmer} M. Wimmer, I. Adagideli, S. Berber, D. Tomanek and K. Richter, Phys. Rev. Lett. {\bf 100}, 177207 (2008)
\bibitem{lundeberg} M. B. Lundeberg, J. A. Folk, Nat. Phys. {\bf 5}, 894 (2009)
\bibitem{Tser} Y. Tserkovnyak, A. Brataas and G. E. W. Bauer, Phys. Rev. Lett. {\bf 88}, 117601 (2002); Phys. Rev. B {\bf 66}, 060404 (2002); Phys. Rev. B {\bf 66}, 224403 (2003); Phys. Rev. B {\bf 67}, 140404 (2003)
\bibitem{bauer-e-cia} Y. Tserkovnyak, A. Brataas, G. E. W. Bauer and B. I. Halperin, Rev. Mod. Physics {\bf 77}, 1375 (2005)
\bibitem{submitted} F. S. M. Guimar\~aes, D. F. Kirwan, A. T. Costa, R. B. Muniz, D. L. Mills and M. S. Ferreira, http://arxiv.org/abs/1002.0731
\bibitem{arkady-prl} A. V. Krasheninnikov, P. O. Lehtinen, A. S. Foster, P. Pyykko, and R. M. Nieminen, Phys. Rev. Lett. {\bf 102}, 126807 (2009)
\bibitem{vene-prb} P. Venezuela, R. B. Muniz, A. T. Costa, D. M. Edwards, S. R. Power and M. S. Ferreira, Phys. Rev. B {\bf 80}, 241413(R) (2009)
\bibitem{gan} Y. Gan, L. Sun, and F. Banhart, Small {\bf 4}, 587 (2008)
\bibitem{ushiro} M. Ushiro, K. Uno, T. Fujikawa, Y. Sato, K. Tohji, F. Watari, W. J. Chun, Y. Koike, and 
K. Asakura, Phys. Rev. B {\bf 73}, 144103 (2006)
\bibitem{jpcm04} O. Cespedes, M. S. Ferreira, S. Sanvito, J. M. D. Coey and M. Kociak, J. Phys.: Condens. Matter {\bf 16}, L155 (2004)
\bibitem{prb04} M. S. Ferreira and S. Sanvito, Phys. Rev. B {\bf 69}, 035407 (2004)
\bibitem{esquinazi} P. Esquinazi, D. Spemann, R. Hšhne, A. Setzer, K.-H. Han, and T. Butz, Phys. Rev. Lett. {\bf 91}, 227201 (2003) 
\bibitem{bauer03} B. Heinrich, Y. Tserkovnyak, G. Woltersdorf, A. Brataas, R. Urban, and G. E. W. Bauer, Phys. Rev. Lett. {\bf 90}, 187601 (2003).
\bibitem{wees} M. V. Costache, S. M. Watts, C. H. van der Wal, and B. J. van Wees, Phys. Rev. B {\bf 78}, 064423 (2008)
\bibitem{charlier-prl04} S. Latil, S. Roche, D. Mayou and J. C. Charlier, Phys. Rev. Lett. {\bf 92}, 256805 (2004)
\bibitem{rocha} C. G. Rocha, A. Wall, A. R. Rocha and M. S. Ferreira, J. Phys.: Cond. Matter {\bf 19}, 346201 (2007)
\bibitem{obs} For simplicity we consider the electronic hopping between carbon atoms to be the same as the hopping between carbon and the magnetic impurity. Within a five-fold degenerate orbital scheme, this consequently requires a large value of U to obtain a reasonable value for the Mn magnetic moment. 
\bibitem{ah-review} A. H. Castro Neto, F. Guinea, N. M. R. Peres, K. S. Novoselov and A. K. Geim, Rev. Mod. Phys. {\bf 81}, 109 (2009) 
\bibitem{coupling1} A. T. Costa, D. F. Kirwan and M. S. Ferreira, Phys. Rev. B {\bf 72}, 085402 (2005)
\bibitem{suppression} D. F. Kirwan, C. G. Rocha, A. T. Costa and M. S. Ferreira, Phys. Rev. B {\bf 77}, 085432 (2008)
\bibitem{carbon} D. F. Kirwan, V. M. de Menezes, C. G. Rocha, A. T. Costa, R. B. Muniz, S. B. Fagan and M. S. Ferreira, Carbon {\bf 47}, 2533 (2009)
\bibitem{rkky} S. Saremi, Phys. Rev. B {\bf 76}, 184430 (2007)



\end{references}
\end{document}